# Theoretical insights into the hydrophobicity of low index CeO$_2$ surfaces

Marco Fronzi[1], M. Hussein N. Assadi[2†], Dorian A. H. Hanaor[3]

[1]International Research Centre for Renewable Energy, State Key Laboratory of Multiphase Flow in Power Engineering, Xi'an Jiaotong University, Xi'an 710049, Shaanxi, China

[2]Center for Green Research on Energy and Environmental Materials (GREEN), National Institute for Materials Science (NIMS), 1-1 Namiki, Tsukuba, Ibaraki 305-0044, Japan

[3]Chair of Advanced Ceramic Materials, Institute for Materials Science and Technology, Technische Universität Berlin, Berlin 10653, Germany

[†]h.assadi2008@ieee.org,

### Abstract

The hydrophobicity of CeO$_2$ surfaces is examined here. Since wettability measurements are extremely sensitive to experimental conditions, we propose a general approach to obtain contact angles between water and ceria surfaces of specified orientations based on density functional calculations. In particular, we analysed the low index surfaces of this oxide to establish their interactions with water. According to our calculations, the CeO$_2$ (111) surface was the most hydrophobic with a contact angle of $\Theta = 112.53°$ followed by (100) with $\Theta = 93.91°$. The CeO$_2$ (110) surface was, on the other hand, mildly hydrophilic with $\Theta = 64.09°$. By combining our calculations with an atomistic thermodynamic approach, we found that the O terminated (100) surface was unstable unless fully covered by molecularly adsorbed water. We also identified a strong attractive interaction between the hydrogen atoms in water molecules and surface oxygen, which gives rise to the hydrophilic behaviour of (110) surfaces. Interestingly, the adsorption of water molecules on the lower-energy (111) surface stabilises oxygen vacancies, which are expected to enhance the catalytic activity of this plane. The findings here shed light on the origin of the intrinsic wettability of rare earth oxides in general and CeO$_2$ surfaces in particular and also explain why CeO$_2$ (100) surface properties are so critically dependant on applied synthesis methods.

## 1. Introduction

Understanding and controlling surface wettability in ceramic materials is paramount towards their implementation in a wide range of high-value novel biomedical and chemical applications. Hydrophobic coatings have been routinely utilised to control the reactivity and mechanical performance of surfaces while hydrophilicity can be used to facilitate microfluidics, self-cleaning behaviour and thermal coatings [1, 2]. Harnessing the intrinsic wettability of unmodified oxide surfaces rather than polymeric coatings or nanostructuring is a valuable pathway towards thermally and chemically robust surface functionality. Furthermore, the wetting characteristics of an oxide surface are indicative of more general surface features, including defect chemistry, energetics and lattice termination, with implications towards surface localised phenomena in general. The wettability of surfaces is generally parametrised in terms of the contact angle of water at the surface ($\Theta$). Typically, materials are defined as super-hydrophilic if $\Theta$ is close to 0°, and super-hydrophobic if $\Theta$ is larger than 150°.

Recent experiments have identified robust intrinsic hydrophobicity of surfaces in rare-earth oxides (REOs), giving a prospect to multiple novel applications [3-6]. Moreover, the wetting characteristics of this class of material are of importance towards interpreting and improving their catalytic activity. In particular, hydrophobicity confers resistance to water deactivation at catalyst surfaces and enhances the adsorption of organic compounds. Consequently, hydrophobicity, or organophilicity, is frequently associated with higher performance and is often a desired trait in applications involving the oxidation of organic compounds and selective synthesis [7-11]. The lower levels of hydroxyl ion driven surface defects associated with hydrophobicity are further known to enhance luminescence [12]. In recent years, interest in hydrophobic surfaces has been further driven by heat transfer applications, as dropwise water condensation has been found to be a highly effective heat transfer mechanism. As intrinsically hydrophobic ceramic materials, REOs are particularly attractive for such applications, relative to surface functionalised materials, owing to their thermal and chemical robustness.

Because of its diverse functionality and the facile tunability of its bulk- and surface properties, cerium oxide, crystallising in a cubic fluorite structure, is the most widely studied REO [13, 14]. $CeO_2$ is often studied in nanoparticle, pristine and doped forms and has been extensively studied for applications in catalysis and biomedicine, in which the material's interaction with adsorbates, in general, and wettability in particular, plays a key role in performance. Studies have shown that adsorption behaviour of various molecular and ionic species diverges significantly between the different crystallographic planes of $CeO_2$, highlighting the importance of application-targeted morphological design to enhance the expression of high-performance surfaces in ceria-based catalysts [15-17].

Hydrophobicity and wetting angles of up to 116° have been reported for unmodified surfaces of $CeO_2$ nanomaterials, which tend to be dominated by (111) plane [12]. Due to its intrinsic hydrophobicity, $CeO_2$ has been proposed for applications in robust anti-icing and wear resistant coatings [18]. Although many experimental investigations have been conducted into the physical properties of this oxide, the relative interactions of its low index surfaces with water remain poorly understood. To gain fundamental mechanistic insights into the behaviour of applied $CeO_2$ surfaces, here we use a theoretical-computational methodology to investigate the hydrophobicity of the principal surfaces of $CeO_2$, namely the terminated (111), (110) and (100) planes, which are dominant in morphologies of octahedra, nanorods and cubes respectively [16, 19]. In the present work we parametrise surface wettability to obtain a measurable quantity that can be compared with experimental results. From the adsorption energies of a water bilayer, we compute the contact angles $\Theta$ to compare the surfaces' hydrophobicity [12].

## 2. Methodology

The calculations presented in this work were performed using unrestricted density functional theory approach, within the generalised gradient approximation (GGA) based on the Perdew-Burke-Ernzerhof (PBE) parameterisation for the exchange-correlation functional [20]. We used periodic boundary conditions and a plane-wave basis set as implemented in VASP code [21-24], where the projector augmented wave pseudopotential is used to replace the core electrons. The cut-off energy for the valence electrons was set at 500 eV. We included a Hubbard correction with a $U_{eff}$(Ce$_{4f}$) of 4.2 eV to improve the electron localisation effects [25]. The inclusion of the Hubbard $U$ term in the DFT+$U$ approach corrects the resulting electronic structure of the reduced CeO$_2$, favouring the insulating state rather than the incorrect metallic solution. However, this approach, in general, suffers from a strong linear dependence of the energetics on the value chosen for the parameter $U$ and also on the localised projector functions chosen that define the $U$-dependent energy term. Here, to circumvent this problem, we verified that the energetics of the systems calculated using $U_{eff}$=4.2 eV agree with the previously calibrated results [26], validating our choice of $U$. Since, in the CeO$_2$ and H$_2$O system, the effect of van der Waals correction is considerably smaller than the dominant localisation effect of $f$ electrons [27], the calculations were limited to the GGA+$U$ framework.

We examined here the three principle low index surfaces of CeO$_2$, corresponding to oxygen terminated (111), (110) and (100) planes. These surfaces were selected due to their significance in the surface-driven functionality of CeO$_2$ across the various applications and growth morphologies of this material [28, 29]. We constructed these surfaces by cleaving the fluorite-type cubic CeO$_2$ structure [30]. The simulated supercells consisted of the cleaved surface slabs with a ($2u \times 2v$) expansion and a depth of three unit cells resulting in a 144 atom supercell for (100), a 96 atoms supercell for (110) and a 60 atom supercell for (111) surfaces. To find the ground state configuration of the adsorbed water layers on the CeO$_2$ surface, we ran quenching *ab initio* molecular dynamics simulations with a target temperature of 20 K with steps of 0.1 fs. The molecular dynamics simulation was effective in finding the initial structures for geometry optimisation. Full geometry optimisation was then carried out on the equilibrated structures, with convergence criteria for the energy and forces of $10^{-5}$ eV and $10^{-2}$ eV/Å, respectively.

# 3.  Results

## 3.1. Hydrophobicity

We calculated the adsorption energy ($E_{Ad}$) of a water double layer composed of eight molecules for (100) and (111) surfaces and sixteen molecules for the (110) surface by subtracting the total DFT energy of the supercell containing both the CeO$_2$ slab and the water double layer from the summed total energies of the individual CeO$_2$ surface slabs and the standalone water double layers. Since the (110) surface has a relatively larger surface cross-section (21.137 Å$^2$) than the (100) and (111) surfaces (14.947 Å$^2$), to fully cover the interface, a larger number of H$_2$O molecules were required. Furthermore, (110) has three undercoordinated surface ions; two O ions and a Ce ion while (100) and (111) surfaces have only two undercoordinated surface ions; two O ions for the (100) surface, and an O ion and a Ce ion for the (111) surface. Consequently, (110) surface required twice as many water molecules to saturate all undercoordinated surface ions. The configurations of the water CeO$_2$ surfaces and the double layers are presented in Figure 1. According to earlier works, the calculated adsorption energy of the water to the CeO$_2$ surface does not significantly change if the thickness of the water slab is increased beyond two layers [31]. Based on this approximation we can use the optimised structures and their total energies calculated here to predict the contact angle between water and CeO$_2$ surfaces.

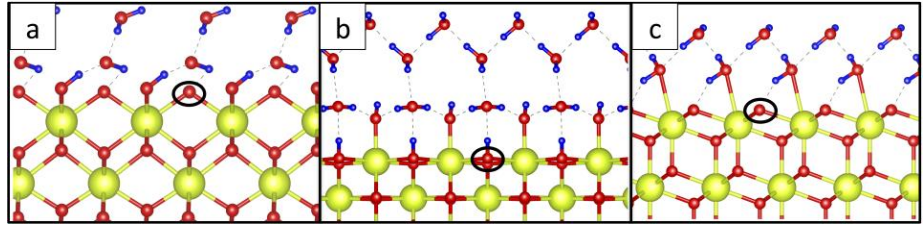

Figure 1. (Color online) Schematic representation of a water double layer adsorption onto CeO$_2$(100) (a), CeO$_2$(110) (b) and CeO$_2$(111) (c) surfaces. Yellow, red and blue spheres represent cerium, oxygen and hydrogen atoms, respectively.

Young's Equation defines the contact angle between a given surface and a liquid, schematically shown in Figure 2(a), as follows [32]:

$$\cos(\theta) = \frac{\gamma_{Solid-Gas} - \gamma_{Solid-Liquid}}{\gamma_{Liquid-Gas}}, \tag{1}$$

where $\gamma$ is the interface energy. In our theoretical framework, these quantities are calculated in the absence of thermal excitation, and therefore we approximate here that the liquid at an atomistic level acts similarly to ice, an approximation justified experimentally by Van Oss *et al*. [33]. Consequently, we consider the interfacial energies with ice rather than liquid water. As a second approximation, we use here vacuum conditions as a substitute for the gas phase, as simulation cells are cleaved by vacuum rather than water vapour. The magnitude of the error resulting from this approximation was shown to be minor by Carchini *et al*. [31]. By applying these two approximations; we obtain a new equation for the contact angle as

$$\cos(\theta) \simeq \frac{\gamma_{CeO_2-vacuum} - \gamma_{CeO_2-ice}}{\gamma_{ice-vacuum}}. \tag{2}$$

The total energy of the supercell containing both the water double layer and ceria surface cell shown in Figure 2(b) can be decomposed as:

$$E_{CeO_2+ice}^{Tot} = A \cdot \gamma_{CeO_2-vacuum} + A \cdot \gamma_{ice-vacuum} + A \cdot \gamma_{CeO_2-ice} + nE_{Ice(Bulk)}^{Tot} + mE_{CeO_2(Bulk)}^{Tot} \tag{3}$$

Here, $E_{CeO_2+ice}^{Tot}$ is the total energy of the supercell containng the CeO$_2$ and adsorbed water, $E_{Ice(Bulk)}^{Tot}$ is the total energy of bulk ice in stable hexagonal form, $E_{CeO_2(Bulk)}^{Tot}$ is the total energy of bulk CeO$_2$ and $\gamma$ are interfacial energies indicated by their respective subscripts. The coeficients $n$ and $m$ correspond to the molecular ratios given by the stoichiometry of the supercell. A schematic representation of interface energies is given in Figure 2. $\gamma_{CeO_2-vacuum}$ can be calculated from the difference of bulk and surface slabs as:

$$2A \cdot \gamma_{CeO_2-vacuum} = E_{CeO_2-slab}^{Tot} - mE_{Bulk-CeO_2}^{Tot}, \tag{4}$$

where $E_{CeO_2-slab}^{Tot}$ is the total energy of the CeO$_2$ slab for a given surface. Similarly, $\gamma_{ice-vacuum}$ , as illustrated in Figure 2(c), can be calculated as

$$2\gamma_{ice-vacuum} = E^{Tot}_{ice-slab} - m E^{Tot}_{Ice(bulk)}. \qquad (5)$$

On the basis of these interfacial energy values, we evaluate Equation (2) and obtain the adsorption energies and contact angles of water with the three surfaces as summarised in Table 1.

**Table 1. Water double layer adsorption energies ($E_{Ad}$) at CeO$_2$ (100), (110) and (111) surfaces calculated with respect to the isolated water double layer (DL), and respective CeO$_2$ slabs are presented in the first two rows. Values are given per unit area and per adsorbed molecule. Furthermore, the calculated $\Theta$ and the distance of Ce ions from their neighbouring Ce ions are at the CeO$_2$/water interface are also given.**

| Configuration | (100) | (110) | (111) |
|---|---|---|---|
| $E_{Ad}$(H2O–DL) (eV/Å$^2$) | −0.023 | −0.038 | −0.014 |
| $E_{Ad}$(H2O–DL) (eV/H$_2$O) | −0.345 | −0.404 | −0.179 |
| $\Theta$ | 93.91° | 64.09° | 112.53° |
| Ce−Ce | 3.86 Å | 3.86, 5.46 Å | 3.86 Å |

The attraction between the water double layer and the CeO$_2$ (110) surface is the strongest followed by the that of the (100) plane, with the (111) surface exhibiting the lowest adsorption energy. These results show that the (111) surface is strongly hydrophobic as its $\Theta$ value substantially exceeds 90°. The (110) surface can, on the other hand, be considered weakly hydrophilic as indicated by its calculated contact angle of $\Theta = 64.09°$ which is smaller than 90°. The (100) surface is slightly hydrophobic as $\Theta$ for this plane is marginally larger than 90°. The value obtained for the (111) surface is slightly larger than the earlier obtained $\Theta$ value of 99.9° calculated by Carchini *et al*. however it is closer to the experimentally obtained value (103°) reported in the same study [31]. To understand the origin of the different levels of hydrophobicity/hydrophilicity of CeO$_2$ surfaces, we analysed the interaction between those hydrogen atoms closest to the CeO$_2$ surface and the surface oxygen atoms by calculating the crystal orbital overlap populations (COOP). COOP is calculated by multiplying overlapping electronic populations by their corresponding density of states. Positive and negative values indicate bonding and antibonding interactions respectively. The COOP plots, presented in Figure 3, show strong bonding of H from the closest adsorbed water layer to the lattice oxygen at the (110) surface as indicated by the peak at ∼−5.5 eV. For the (100) surface, the interaction of H with O results in more moderate bonding states at ∼−2 eV and ∼−7 eV. At the (111) surface, on the other hand, there are only minor bonding states close to the Fermi level with peaks at ∼−1 eV with no strong bonding peaks anywhere deeper in the valence band. We thus observe here a correlation between the contact angle and the intensity of the bonding states, where more intense H−O bonding peaks deeper in the valence band indicate higher hydrophilic water ceria interaction. Indeed, the (110) surface has the most intense H−O bonding peak and the (111) the least intense, in line with the calculated contact angles that indicate a hydrophilic and hydrophobic behaviour for the (110) and the (111) surfaces, respectively. Furthermore, the O−H interaction at (100) and (110) surfaces results in antibonding states at ∼−1 eV indicating that the surface oxygens dissociate the water molecules and form separated H and H−O groups. Water dissociation within the first layer of water is evident from the relaxed structures presented in Figure 1(a) and (b).

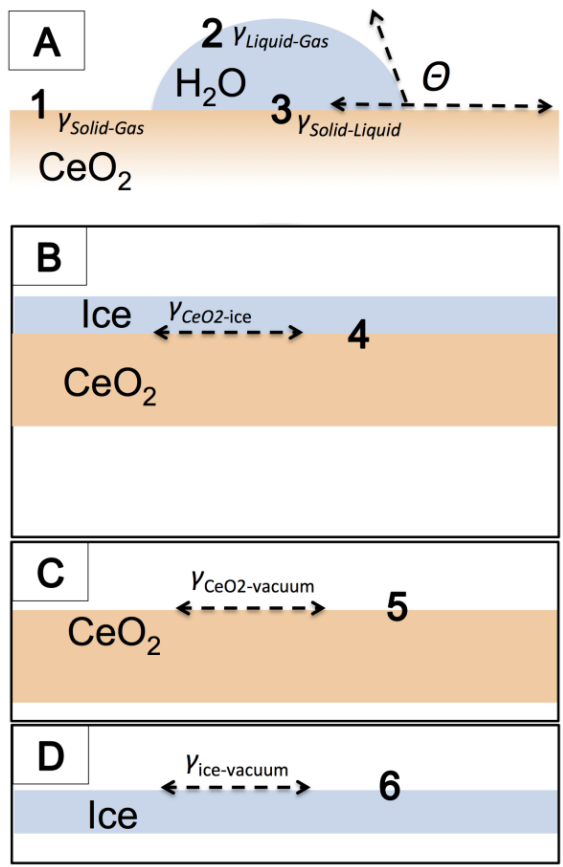

**Figure 2. Schematic representation of A) water adsorption on a CeO₂ surface B), C) and D) the models used in our calculations. Here we indicate the interfaces used in Equations 2-5, where each number indicates a specific interface as follow:1 the solid-gas, 2 the liquid-gas, 3 the solid-liquid, 4 the CeO₂-ice, 5 the CeO₂-vacuum, 6 ice-vacuum.**

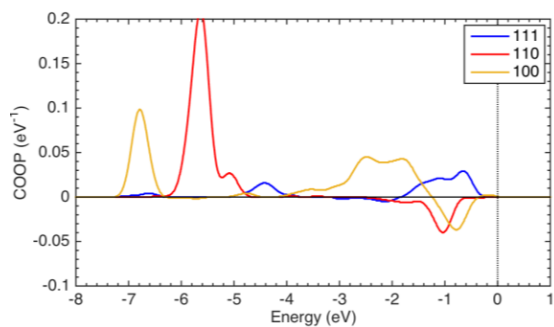

**Figure 3. (Color online) COOP of CeO₂ (100) (yellow line), CeO₂ (110) (red line) and CeO₂ (111) (blue line) corresponding to the closest H from the water double layer to the CeO₂ surface and the outermost O from CeO₂ surface. The vertical line at 0 eV represents the Fermi level energy.**

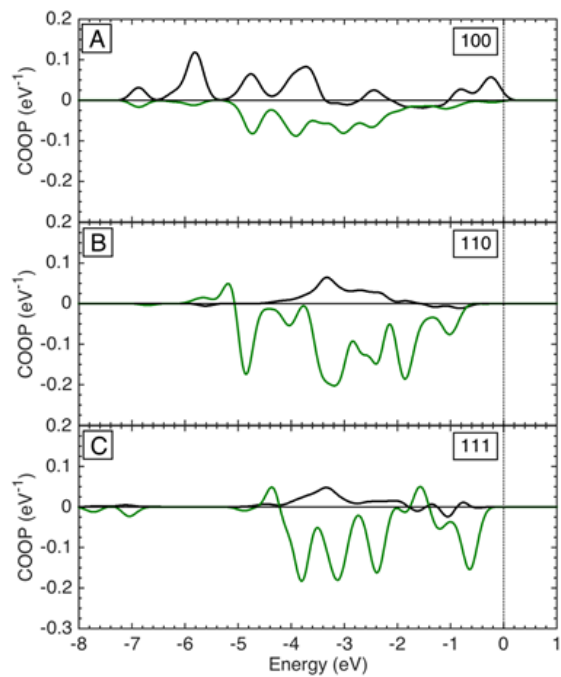

**Figure 4. (Color online) COOP of a) CeO₂ (100), b) CeO₂ (110) and c) CeO₂ (111) surfaces upon water double layer adsorption. The black line represents the COOP between the surface cerium and surface oxygen atoms. The green line represents COOP between the surface cerium atoms and two nearest surface cerium atoms. The vertical line at 0 eV represents the Fermi level energy.**

To analyse the nature of bonding in the CeO₂ outermost layer at the water/CeO₂ interface, we calculated the COOP corresponding the overlap between neighbouring surface Ce–Ce and Ce–O pairs in the CeO₂ slab as presented in Figure 4. Accordingly, the Ce–Ce COOP plots for all surfaces are dominated by antibonding states for energies between ∼−5 eV and ∼−2 eV. Additionally, there are two minor bonding Ce–Ce peaks for (111) at ∼−1.5 eV and ∼−4.3 eV. For the (110) surface, there is one minor bonding region at −5.0 eV. The Ce–Ce COOP for the (100) surface, however, shows no bonding states. Given that the Ce–Ce interaction is dominated by antibonding states, surface stability must, therefore, be driven by Ce–O or O–H interactions. The Ce–O interactions at the (110) and (111) surfaces generally show strong bonding character for deeper energies of the valence band (E <∼−1.5 eV). The (100) surface, on the other hand, has two broad antibonding states at E = ∼−1.5 eV and E = ∼−3 eV that suggest the instability of this termination in vacuum conditions in the absence of O–H bonding states when fully covered with the H₂O layer as shown in Figure 3.

## 3.2. Surface Thermodynamic Analysis

As Ce⁴⁺ can be readily reduced to Ce³⁺ in CeO₂ alongside the liberation of O, the effects of such vacancies merit consideration and have been partially examined experimentally by high-resolution scanning force microscopy [34]. Here we calculated the formation energy of an oxygen vacancy at the (100), (110) and (111) surfaces according to the following equation:

$$E_{V_O} = E_{Surf+V_O}^{Tot} - E_{Surf}^{Tot} + \frac{1}{2}E_{O_2}^{Tot}, \qquad\qquad (6)$$

where $E_{Surf+V_O}^{Tot}$ is the total energy of the surface following vacancy formation (reduced surface), $E_{Surf}^{Tot}$ is the total energy of the pristine (stoichiometric) surface and $E_{O_2}^{Tot}$ is the total energy of an oxygen molecule calculated in a vacuum box. The results are presented in the first row of Table 2. The formation energy of $V_O$ on the (100) surface has a negative value which indicates that bare O terminated (100) surface is not stable without water adsorption, as implied also by the COOP analysis of the preceding section. Specifically, given the two broad antibonding Ce–O COOP peaks in (100) shown in Figure 4(a), a negative formation energy for $V_O$ is not surprising at this surface. However, the formation energy of vacating the same O site at the (100) surface covered with the water double layer, marked with a circle in Figure 1(a), is 0.608 eV, leading to the conclusion that O terminated (100) surface is stablised in aqueous conditions. The (110) and (111) surfaces, on the other hand, are stable under any conditions of water adsorption as the $V_O$ formation energy has a positive values on these surfaces.

As O terminated (100) surface cannot be stabilised by partial $H_2O$ covering, we focus on the interaction of a single water molecule with (110) and (111) surfaces under the stoichiometric and reduced conditions to understand how these surfaces behave when partially covered with water. Water adsorption energies were calculated according to the following equation:

$$E_{Ads} = E_{Surf+H_2O}^{Tot} - E_{Surf}^{Tot} - E_{H_2O}^{Tot}, \qquad\qquad (7)$$

where $E_{Surf+H_2O}^{Tot}$ is the total energy of the surface upon water adsorption, $E_{Surf}^{Tot}$ is the total energy of the initial surface which can be either stoichiometric or reduced, and $E_{H_2O}^{Tot}$ is the total energy of a water molecule calculated in a vacuum box. Furthermore, for every scenario, two different configurations were considered; first molecular (or intact) adsorption of the water molecule in which an intact water molecule is adsorbed at a Ce site and second, a dissociated water molecule which consists of a HO group adsorbed on Ce site and a H atom adsorbed on a nearby O site. The relaxed structures of all configurations are presented in Figure 5.

As shown in Table 2, by comparing the molecular water adsorption on both surfaces, we observe that $E_{Ads}(H_2O)$ is lower at the (110) surface. Furthermore, the formation energy of an O vacancy was significantly larger at $CeO_2$ (111) surface than that of (110) surface. Higher $V_O$ formation energy, in turn, results in lower molecular and dissociative water adsorption energies onto the reduced (111) surface. That is because bonding between the reduced (111) surface with the O atom of the adsorbed water molecule compensates for the chemical instability that was created by the removal of an oxygen atom at the (111) surface. The Results obtained here are in general agreement with the earlier GGA calculations [26].

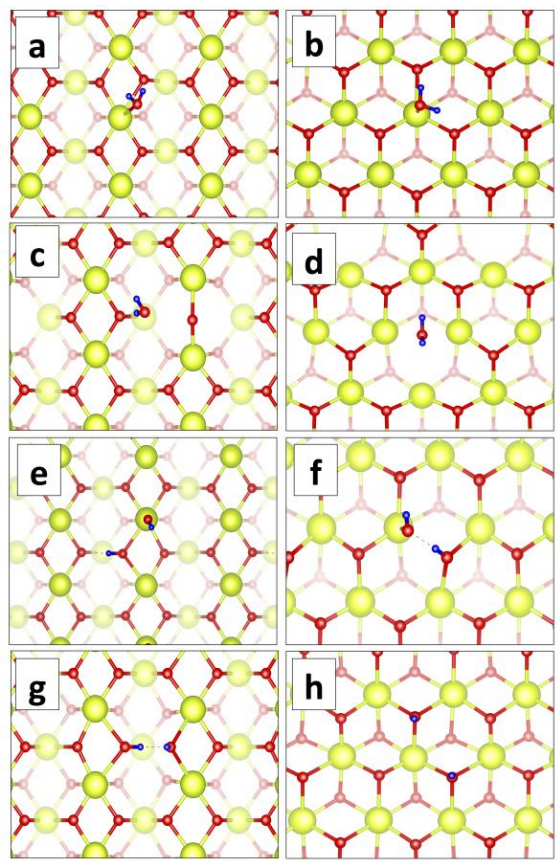

**Figure 5.** (Color online) Schematic representation of molecular and dissociative water adsorption onto CeO₂(110) (left) and (111) (right) surfaces. From top to bottom, molecular adsorption onto stoichiometric surfaces (a) and (b), molecular adsorption onto reduced surfaces (c) and (d), dissociative adsorption onto stoichiometric surfaces (e) and (f) and dissociative adsorption onto reduced surfaces (g) and (h). Yellow, red and blue spheres represent cerium, oxygen and hydrogen atoms, respectively.

**Table 2.** Vacancy formation as well as water molecular (associative) and dissociative adsorption energies at CeO₂ (111) and CeO₂ (110) surfaces (calculated with respect to the water in the gas phase), schematically represented in Fig.5. Adsorption energies were not calculated for (100) surface due to the lack of stability of O termination in this surface when not fully covered with H₂O.

| Configuration | (100) | (110) | (111) |
|---|---|---|---|
| $V_O$ | −3.784 | 1.56 eV | 2.72 eV |
| $H_2O$ | - | −0.64 eV | −0.54 eV |
| OH–H | - | −1.11 eV | −0.55 eV |
| $V_O$–$H_2O$ | - | −0.25 eV | −0.77 eV |
| $V_O$–OH–H | - | −1.17 eV | −2.49 eV |

We then calculated surface energies, within the framework of the *ab-initio* thermodynamics, with the inclusion of the chemical potentials of water and oxygen [26, 35]. We considered chemical potentials as independent variables so that a wider range of conditions could be extrapolated. Assuming that the gas reservoir exchanges particles with the surface without affecting its overall chemical potential, the surface energy can be calculated as follows:

$$\gamma(p,T) = \frac{1}{A}[G - \sum N_i \mu_i(p,T)], \tag{8}$$

where $A$ is the interface area, $G$ is the Gibbs free energy of the crystal, $\sum N_i \mu_i(p,T)$ is the chemical potential of the atomic species in the system and the term $N_i$ is the total number of atoms of the species $i$. We assume that the interface is in equilibrium with an environment formed by a mixture of two different chemical species in the gas-phase, namely $O_2$ and $H_2O$, which are considered to be in non-equilibrium with one another. We measure the chemical potentials with respect to the calculated energy of the isolated molecules, and thus define the relative chemical potentials to be $\Delta\mu_i(p,T) = \mu_i - E_i$. By expanding Equation 8 to obtain a relationship between $\gamma(p,T)$, $\gamma_O(p,T)$ and $\gamma_{H_2O}(p,T)$, where water and oxygen chemical potential are treated as independent variables, we obtain phase diagrmas of stable surface configurations as shown in Figure 6 in which each colored area indicates which surface termination is stable with respect to the other competing surfaces for any given values of chemical potentials. The borders of different areas indicate the transitions between the stability of two surface terminations, and give us information on the conditions for such transitions. For CeO$_2$(100) we obtain that water termination is stable under any environmental conditions, confirming the conclusions reached in the previous section. We therefore do not include this phase diagram in Figure 6, since it would not bring any useful information.

Furthermore, we observe that for the (111) surface there is no direct transition between the region corresponding the stability of water adsorbed at the stoichiometric surface and that corresponding to the bare reduced surface (denoted by 111-V$_O$). For the (110) surface there is no direct transition between the stability field of the pristine surface and that of water terminated reduced surface. These phase diagrams thus demonstrate that for the (111) surface the presence of water stabilises surface oxygen vacancies, shifting the stability of V$_O$ towards higher values of the oxygen chemical potential. This is shown in the top panel of Figure 6 where H$_2$O–V$_O$, water adsorbed on a reduced surface, is stable for higher $\Delta\mu_O$ compared to the water adsorbed on the bare reduced surface that contains V$_O$. This point can be utilised experimentally as the stability of oxygen vacancy results in high chemical activity for ceria surface through the Mars-van Krevelen mechanism in which redox reactions are mediated by oxygen vacancies [36]. Furthermore, on reduced ceria surfaces, superoxides could, in principle, be formed when an electron, trapped at a Ce$^{3+}$ site, is transferred to an adsorbed O$_2$ molecule which is a critical mechanism in a wide variety of reaction pathways [37]. Unlike the (111) surface, for CeO$_2$ (110), the presence of water molecules does not stabilise surface V$_O$, and therefore no increase in surface chemical activity is anticipated in the presence of high H$_2$O chemical potential.

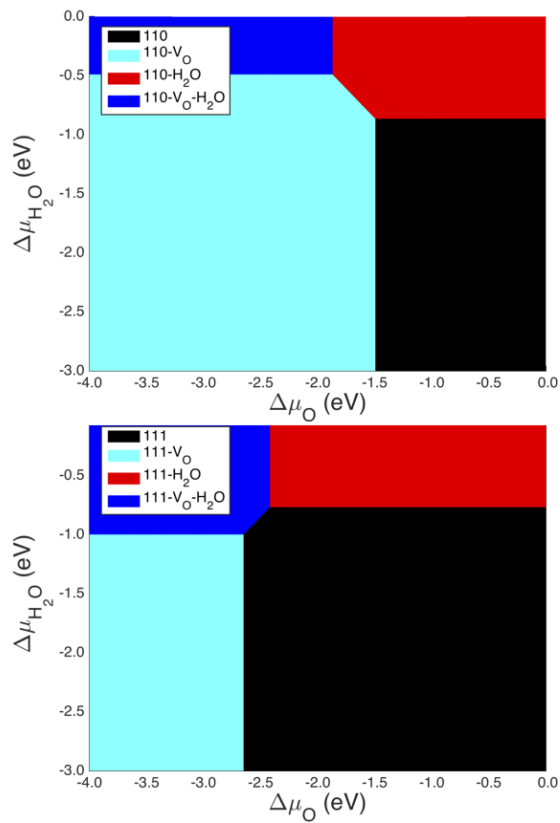

**Figure 6. (Color online) Surface phase diagram of stable structures of CeO₂(111) and CeO₂(110) in equilibrium with an oxidising and "humid environment", as a function of $\Delta\mu_O$ and $\Delta\mu_{H_2O}$ in the gas phase.**

Finally, we revisited the contact angles taking into account the presence of $V_O$ at CeO₂ and H₂O interface, marked with circles in Figure 1. We repeated the calculations of Section 3.1 using $4u\times4v$ extended supercells for which one $V_O$ was introduced to all interfaces. The $V_O$ concentration in the extended supercell was 3.125% for (100), (110) surfaces and 6.25% for the (111) surface. The contact angles for the O deficient surfaces was calculated to be 118.59° for (100) surface, 74.30° for (110) surface and 160.15° for (111) surface. Although we anticipate that $\Theta$ is sensitive to the $V_O$ concentration, these calculations, nonetheless, demonstrate the O vacancy on the low index CeO₂ surfaces, in general, increases the hydrophobicity.

# 4. Discussion

Numerous rare earth oxides have been found to exhibit intrinsic hydrophobicity, i.e. without the use of chemical surface treatments. Of these REOs, one can consider CeO₂ to be the most attractive, due to its relative availability and diverse functionality in different forms and microstructures. The origins of hydrophobicity in CeO₂ have been the subject of discussions in recent literature [3, 38]. By density functional theory (DFT) and crystal orbital overlap population (COOP), we have demonstrated here that the hydrophilicity or hydrophobicity of the different surfaces of rare earth oxides, such as $CeO_2$, can be understood in light of their electronic and crystallographic structures and the consequent bonding states. The weaker H–O bonding at the $CeO_2(111)$ surface in the form of minor shallow bonding states indicated in COOP analysis, corresponds to a highly hydrophobic surface, consistent with the numerous experimental studies on this material and confirming the large contact angle calculated here [39, 40]. The deeper bonding states found in (100) and strong bonding states for (110) correspond to the greater wettability of these planes.

The dissociative adsorption of water at the (111) and (100) surfaces has been observed experimentally, with analysis showing dissociation is relatively more favoured on (100) surfaces, with dissociation being enhanced for both cases in reduced materials [41, 42], in qualitative agreement with the results of the present study. Further studies, by atom force microscopy, have also shown the dissociation of water to be driven by the presence of oxygen vacancies at surface sites on the lower energy (111) planes, as hydroxide substitutes for surface oxygen [34].

Numerous experimental studies, mainly related to CO oxidation, have observed poorer catalytic activities at (111) planes in $CeO_2$ relative to the higher energy (100) and (110) planes [43]. In contrast, as a support for catalytic nanoparticles (Au, Cu, Pt) in water gas shift reactions, (111) plane of ceria is found to impart enhanced performance [44, 45]. Whether wettability of a surface contributes or detracts from a heterogeneous catalyst activity depends on diverse mechanistic aspects, and therefore the design of $CeO_2$ structures remains non-trivial. However, the findings here, indicating the stabilisation of surface oxygen vacancies in the presence of molecularly adsorbed water on (111) surfaces, suggesting that for catalytic pathways in aqueous systems involving vacancy mediated redox reactions, the increased prominence of this plane (such as in octahedral nanocrystals) may be advantageous.

According to our findings, the negative value of the O vacancy formation energy on the polar (100) surface demonstrates that this surface is intrinsically unstable. Experimentally, various stabilising surface reconstructions have been reported that are highly dependent on the sample preparation and treatment history. Capdevila-Cortada et al. showed that surface disorder created by partial removal of O could kinetically stabilise the bare (100) surface through fast surface diffusion process among nearly degenerate $V_O$ configurations.[46]. Judging from the positive value of the $V_O$ formation energy on fully water covered (100) surface, we infer that full water coverage can also provide a stabilisation mechanism for $CeO_2$ (100) surface.

It has also been argued that a larger average Ce–Ce distance at the interface results in higher levels of hydrophilicity, implying that the Ce geometry at the interface is the main driving force behind the wettability of the system [31]. In contrast, as shown in Table 1, in the most hydrophobic surface i.e. the (111) surface, the Ce atoms have six Ce nearest neighbours with a Ce−Ce distance of 3.86 Å, whereas for the most hydrophilic (110) surface, Ce has only two nearest neighbours with Ce−Ce distance of 3.86 and two second nearest neighbours at 5.46 Å. Also, in (100) surface Ce has only four nearest neighbours with a distance of 3.86 Å and four second nearest neighbours with a distance of 5.47 Å. Consequently, our results suggest that large Ce−Ce distances are positively correlated with hydrophobicity rather than hydrophilicity. Therefore the previously stated correlation between Ce−Ce distances and hydrophobicity does not constitute a general trend, and instead, a positive correlation can be established only through the analysis of the specific electronic configuration and COOP [31].

## 5. Conclusions

$CeO_2$ (111) surface was found to be the most hydrophobic with $\Theta = 112.53°$ followed by (100) with $\Theta = 93.91°$. In contrast, the $CeO_2$ (110) surface, was found the most hydrophilic with $\Theta = 64.09°$. The contact angles increase for all three surfaces when O vacancy is present. Moreover, the O terminated (100) surface was found to be unstable in the absence of full coverage by molecular $H_2O$. In the absence of water, $CeO_2$ (100) surfaces will be, at least, partially Ce terminated. This finding implies that $CeO_2$ (100) surfaces formed in an aqueous environment (through sol-gel synthesis for example) are likely to exhibit different surface properties from materials prepared using vacuum based methods (e.g. pulsed laser deposition). Finally, the thermodynamic analysis demonstrates that water adsorption stabilises oxygen vacancies on $CeO_2$ (111) surfaces, shifting the stability of $V_O$ towards higher values of the oxygen chemical potential. The increased concentration of $V_O$ at higher oxygen chemical potentials is expected to improve the catalytic activity of (111) surface, while the opposite trend was found for $CeO_2$ (110).

## Acknowledgements

The financial support was provided by the National Natural Science Foundation of China (Grant No. 51323011). Computational resources were provided by the Center for Computational Sciences at the University of Tsukuba.

## References

[1] P.F. Rios, H. Dodiuk, S. Kenig, S. McCarthy, A. Dotan, Polym. Adv. Technol., Durable ultra-hydrophobic surfaces for self-cleaning applications. 19 (2008) 1684–1691.

[2] B. Peng, X. Ma, Z. Lan, W. Xu, R. Wen, Int. J. Heat Mass Transfer, Experimental investigation on steam condensation heat transfer enhancement with vertically patterned hydrophobic–hydrophilic hybrid surfaces. 83 (2015) 27–38.

[3] G. Azimi, R. Dhiman, H.-M. Kwon, A.T. Paxson, K.K. Varanasi, Nat. Mater., Hydrophobicity of rare-earth oxide ceramics. 12 (2013) 315.

[4] Y. Tian, L. Jiang, Nat. Mater., Intrinsically robust hydrophobicity. 12 (2013) 291–292.

[5] I.-K. Oh, K. Kim, Z. Lee, K.Y. Ko, C.-W. Lee, S.J. Lee, J.M. Myung, C. Lansalot-Matras, W. Noh, C. Dussarrat, H. Kim, H.-B.-R. Lee, Chem. Mater., Hydrophobicity of rare earth oxides grown by atomic layer deposition. 27 (2015) 148–156.

[6] C. Reitz, J. Haetge, C. Suchomski, T. Brezesinski, Chem. Mater., Facile and general Synthesis of thermally stable ordered mesoporous rare-earth oxide ceramic thin films with uniform mid-size to large-size pores and strong crystalline texture. 25 (2013) 4633–4642.

[7] E. Nowak, G. Combes, E.H. Stitt, A.W. Pacek, Powder Technol., A comparison of contact angle measurement techniques applied to highly porous catalyst supports. 233 (2013) 52–64.

[8] K.T. Chuang, B. Zhou, S. Tong, Ind. Eng. Chem. Res., Kinetics and mechanism of catalytic oxidation of formaldehyde over hydrophobic catalysts. 33 (1994) 1680–1686.

[9] J. Chi-Sheng Wu, T.-Y. Chang, Catal. Today, VOC deep oxidation over Pt catalysts using hydrophobic supports. 44 (1998) 111–118.

[10] J.-P. Dacquin, H.E. Cross, D.R. Brown, T. Duren, J.J. Williams, A.F. Lee, K. Wilson, Green Chem., Interdependent lateral interactions, hydrophobicity and acid strength and their influence on the catalytic activity of nanoporous sulfonic acid silicas. 12 (2010) 1383–1391.

[11] G. Francesco, G. Michelangelo, M.P. Lo, R. Serena, N. Renato, Adv. Synth. Catal., New Simple Hydrophobic Proline Derivatives as Highly Active and stereoselective catalysts for the direct asymmetric Aldol reaction in aqueous medium. 350 (2008) 2747–2760.

[12] X. Zheng, L. Liu, X. Zhou, Colloid J., Formation and properties of hydrophobic $CeO_2$ nanoparticles. 76 (2014) 558–563.

[13] A. Trovarelli, Catal. Rev., Catalytic properties of ceria and $CeO_2$-containing materials. 38 (1996) 439–520.

[14] B. Choudhury, P. Chetri, A. Choudhury, J. Exp. Nanosci., Annealing temperature and oxygen-vacancy-dependent variation of lattice strain, band gap and luminescence properties of $CeO_2$ nanoparticles. 10 (2015) 103–114.

[15] W. Pengren, P. Chaoyi, W. Binrui, Y. Zhiqing, Y. Fubiao, Z. Jingcheng, IOP Conf. Ser.: Mater. Sci. Eng., A facile method of fabricating mechanical durable anti-icing coatings based on $CeO_2$ microparticles. 87 (2015) 012062.

[16] L. Huang, W. Zhang, K. Chen, W. Zhu, X. Liu, R. Wang, X. Zhang, N. Hu, Y. Suo, J. Wang, Chem. Eng. J., Facet-selective response of trigger molecule to $CeO_2$ {110} for up-regulating oxidase-like activity. 330 (2017) 746–752.

[17] J. Tam, U. Erb, G. Azimi, MRS Adv., Non-wetting nickel-cerium oxide composite coatings with remarkable wear stability. 3 (2018) 1647–1651.

[18] S. Agarwal, L. Lefferts, B.L. Mojet, D.A.J.M. Ligthart, E.J.M. Hensen, D.R.G. Mitchell, W.J. Erasmus, B.G. Anderson, E.J. Olivier, J.H. Neethling, A.K. Datye, ChemSusChem, Exposed surfaces on shape-controlled ceria nanoparticles revealed through AC-TEM and water–gas shift reactivity. 6 (2013) 1898–1906.

[19] L. Yan, R. Yu, J. Chen, X. Xing, Cryst. Growth Des., Template-free hydrothermal synthesis of $CeO_2$ nano-octahedrons and nanorods: Investigation of the morphology evolution. 8 (2008) 1474–1477.

[20] J.P. Perdew, K. Burke, M. Ernzerhof, Phys. Rev. Lett., Generalized gradient approximation made simple. 77 (1996) 3865–3868.

[21] G. Kresse, J. Hafner, J. Phys.: Condens. Matter, Norm-conserving and ultrasoft pseudopotentials for first-row and transition elements. 6 (1994) 8245.

[22] G. Kresse, D. Joubert, Phys. Rev. B, From ultrasoft pseudopotentials to the projector augmented-wave method. 59 (1999) 1758–1775.

[23] P. Raybaud, G. Kresse, J. Hafner, H. Toulhoat, J. Phys.: Condens. Matter, Ab initio density functional studies of transition-metal sulphides: I. Crystal structure and cohesive properties. 9 (1997) 11085.

[24] D.R. Hamann, M. Schlüter, C. Chiang, Phys. Rev. Lett., Norm-conserving pseudopotentials. 43 (1979) 1494–1497.

[25] M.H.N. Assadi, D.A.H. Hanaor, J. Appl. Phys., Theoretical study on copper's energetics and magnetism in $TiO_2$ polymorphs. 113 (2013) 233913.

[26] M. Fronzi, S. Piccinin, B. Delley, E. Traversa, C. Stampfl, Phys. Chem. Chem. Phys., Water adsorption on the stoichiometric and reduced $CeO_2$(111) surface: a first-principles investigation. 11 (2009) 9188–9199.

[27] D. Fernandez-Torre, K. Kośmider, J. Carrasco, M.V.n. Ganduglia-Pirovano, R.n. Pérez, J. Phys. Chem. C, Insight into the adsorption of water on the clean $CeO_2$ (111) surface with van der Waals and hybrid density functionals. 116 (2012) 13584–13593.

[28] Z. Yang, T.K. Woo, K. Hermansson, Chem. Phys. Lett., Strong and weak adsorption of CO on $CeO_2$ surfaces from first principles calculations. 396 (2004) 384–392.

[29] T. Désaunay, G. Bonura, V. Chiodo, S. Freni, J.P. Couziniè, J. Bourgon, A. Ringuedé, F. Labat, C. Adamo, M. Cassir, J. Catal., Surface-dependent oxidation of $H_2$ on $CeO_2$ surfaces. 297 (2013) 193–201.

[30] C. Loschen, J. Carrasco, K.M. Neyman, F. Illas, Phys. Rev. B, First-principles LDA+U and GGA+U study of cerium oxides: Dependence on the effective U parameter. 75 (2007) 035115.

[31] G. Carchini, M. García-Melchor, Z. Łodziana, N. López, ACS Appl. Mater. Interfaces, Understanding and Tuning the Intrinsic Hydrophobicity of rare-earth oxides: A DFT+U Study. 8 (2016) 152–160.

[32] T. Young, Philos. Trans. R. Soc. London, III. An essay on the cohesion of fluids. 95 (1805) 65–87.

[33] C.J. Van Oss, R.F. Giese, Z. Li, K. Murphy, J. Norris, M.K. Chaudhury, R.J. Good, J. Adhes. Sci. Technol., Determination of contact angles and pore sizes of porous media by column and thin layer wicking. 6 (1992) 413–428.

[34] S. Gritschneder, M. Reichling, Nanotechnology, Structural elements of $CeO_2$(111) surfaces. 18 (2007) 044024.

[35] M. Fronzi, S. Piccinin, B. Delley, E. Traversa, C. Stampfl, RSC Adv., $CH_x$ adsorption (x = 1–4) and thermodynamic stability on the $CeO_2$(111) surface: a first-principles investigation. 4 (2014) 12245–12251.

[36] C. Doornkamp, V. Ponec, J. Mol. Catal. A: Chem., The universal character of the Mars and Van Krevelen mechanism. 162 (2000) 19–32.

[37] M. Hayyan, M.A. Hashim, I.M. AlNashef, Chem. Rev., Superoxide Ion: Generation and Chemical Implications. 116 (2016) 3029–3085.

[38] T. An, X. Deng, S. Liu, S. Wang, J. Ju, C. Dou, Ceram. Int., Growth and roughness dependent wetting properties of $CeO_2$ films prepared by glancing angle deposition. 44 (2018) 9742–9745.

[39] M.S. Kabir, P. Munroe, V. Gonçales, Z. Zhou, Z. Xie, Surf. Coat. Technol., Structure and properties of hydrophobic $CeO_{2-x}$ coatings synthesized by reactive magnetron sputtering for biomedical applications. 349 (2018) 667–676.

[40] T. Kropp, J. Paier, J. Sauer, J. Phys. Chem. C, Interactions of water with the (111) and (100) surfaces of ceria. 121 (2017) 21571–21578.

[41] D.R. Mullins, P.M. Albrecht, T.L. Chen, F.C. Calaza, M.D. Biegalski, H.M. Christen, S.H. Overbury, J. Phys. Chem. C, Water dissociation on $CeO_2(100)$ and $CeO_2(111)$ thin films. 116 (2012) 19419–19428.

[42] B. Chen, Y. Ma, L. Ding, L. Xu, Z. Wu, Q. Yuan, W. Huang, J. Phys. Chem. C, Reactivity of hydroxyls and water on a $CeO_2(111)$ thin film surface: The role of oxygen vacancy. 117 (2013) 5800–5810.

[43] Tana, M. Zhang, J. Li, H. Li, Y. Li, W. Shen, Catal. Today, Morphology-dependent redox and catalytic properties of $CeO_2$ nanostructures: Nanowires, nanorods and nanoparticles. 148 (2009) 179–183.

[44] J.A. Rodriguez, P. Liu, J. Hrbek, J. Evans, M. Pérez, Angew. Chem. Int. Ed., Water gas shift reaction on Cu and Au nanoparticles supported on $CeO_2(111)$ and $ZnO(000-1)$: Intrinsic activity and importance of support interactions. 46 (2007) 1329–1332.

[45] M. Baron, O. Bondarchuk, D. Stacchiola, S. Shaikhutdinov, H.J. Freund, J. Phys. Chem. C, Interaction of gold with cerium oxide supports: $CeO_2(111)$ thin films vs $CeO_x$ Nanoparticles. 113 (2009) 6042–6049.

[46] M. Capdevila-Cortada, N. López, Nat. Mater., Entropic contributions enhance polarity compensation for $CeO_2$ (100) surfaces. 16 (2016) 328–334.